\documentclass[12pt]{iopart}
\usepackage{iopams}
\usepackage{epsfig}
\usepackage{amssymb}
\usepackage{amstext}
\newtheorem{theorem}{\indent Theorem}

\renewenvironment{pmatrix}
    {\left(\!\begin{array}{*{20}c}}%
    {\end{array}\!\right)}
\renewenvironment{cases}
    {\left\{\!\begin{array}{ll}}%
    {\end{array}\right.}

\begin{document}

\title[Dong and Petersen, SMC for Focus on Quantum Control]{Sliding mode control of quantum systems}

\author{Daoyi Dong}

\address{School of Information Technology
and Electrical Engineering, University of New South Wales at the
Australian Defence Force Academy, Canberra, ACT 2600, Australia.}
\ead{daoyidong@gmail.com}

\author{Ian R. Petersen}

\address{School of Information Technology
and Electrical Engineering, University of New South Wales at the
Australian Defence Force Academy, Canberra, ACT 2600, Australia.}
\ead{i.r.petersen@gmail.com}

\begin{abstract}
This paper proposes a new robust control method for quantum
systems with uncertainties involving sliding mode control (SMC).
Sliding mode control is a widely used approach in classical
control theory and industrial applications. We show that SMC is
also a useful method for robust control of quantum systems. In
this paper, we define two specific classes of sliding modes (i.e.,
eigenstates and state subspaces) and propose two novel methods
combining unitary control and periodic projective measurements for
the design of quantum sliding mode control systems. Two examples
including a two-level system and a three-level system are
presented to demonstrate the proposed SMC method. One of main
features of the proposed method is that the designed control laws
can guarantee desired control performance in the presence of
uncertainties in the system Hamiltonian. This sliding mode control
approach provides a useful control theoretic tool for robust
quantum information processing with uncertainties.

\end{abstract}

\maketitle

\section{Introduction}
Controlling quantum phenomena has been an implicit goal even since
the establishment of quantum mechanics \cite{Warren et al
1993}-\cite{Chu 2002}. Many practical tasks arising from atomic
physics \cite{Chu 2002}, molecular chemistry \cite{Rabitz et al
2000}, \cite{Rice and Zhao 2000}-\cite{Shapiro and Brumer 2003} and
quantum optics \cite{Dowling and Milburn 2003} can be formulated as
quantum control problems. It is desired to develop quantum control
theory to establish a firm theoretic footing for the active control
of quantum systems. Quantum control theory has played an important
role in guiding the control of chemical reactions \cite{Rabitz et al
2000}, \cite{Shapiro and Brumer 2003}. Recently, the development of
quantum control theory has been recognized as a key task required to
establish practical quantum information technology \cite{Dowling and
Milburn 2003}, \cite{Nielsen and Chuang 2000}, \cite{Santos and
Viola 2008}, \cite{Viola and Lloyd 1998}. Some useful tools from
classical control theory, such as optimal control theory
\cite{Peirce et al 1988}, \cite{Khaneja et al 2001} and feedback
control approaches \cite{Wiseman and Milburn 1993}-\cite{Yamamoto et
al 2008} have been applied to problems of population transfer,
quantum state preparation, quantum error correction and decoherence
control \cite{Branderhorst et al 2008}.

Although some useful results have already been obtained, research on
quantum control is still in its infancy. From the perspective of
practical applications, it is inevitable that there exist
uncertainties and disturbances in external fields, system
Hamiltonians or initial states \cite{Brown and Rabitz 2002}. Many
cases of unknown information and errors can also be treated as
uncertainties. Hence, the requirement of a certain degree of
robustness in the presence of uncertainties and noises has been
recognized as one of the key aspects for developing practical
quantum technologies \cite{Pravia et al 2003}-\cite{D'Helon and
James 2006}. Several approaches have been introduced to enhance the
robustness of quantum systems. For example, Zhang and Rabitz
\cite{Zhang and Rabitz 1994} used a minmax approach to analyze the
robustness of molecular systems. In \cite{James et al 2008}, James
and coworkers have formulated and solved a quantum robust control
problem using a $H^{\infty}$ method for linear quantum stochastic
systems. In this paper, we develop a new sliding mode control
approach to enhance the robustness of quantum systems. Sliding mode
control (SMC) is a useful robust control strategy in classical
control theory and industrial applications \cite{Utkin
1977}-\cite{Utkin et al 1999}. References \cite{Dong and Petersen
2008} and \cite{Vilela Mendes and Man'ko 2003} have briefly
discussed the possible application of sliding mode control to
quantum systems. This paper will formally present sliding mode
control methods for quantum systems to deal with uncertainties.

The sliding mode control approach generally includes two main steps:
selecting a sliding surface (sliding mode) and controlling the
system state to and maintaining it in this sliding surface. The
sliding surface guarantees that the quantum system has desired
dynamics in this surface. We will select an eigenstate of the free
Hamiltonian or a state subspace of the controlled quantum system as
a sliding mode. To control the system state to and then maintain it
in this sliding surface, most feedback control methods in classical
control theory are not directly applicable since we generally cannot
acquire measurement feedback information without destroying the
quantum system state. It is necessary to develop new approaches to
accomplish this task. Hence, we propose two new unique methods for
this task: one method involves combining time-optimal control design
and periodic projective measurements for the case where an
eigenstate is taken as the sliding mode; the other method is
implemented through quantum amplitude amplification and periodic
projective measurements for the case where a state subspace is
identified as a sliding mode. It is worth noting that an important
assumption required in the proposed methods is that accurate
projective measurements (using the eigenstates of the free
Hamiltonian as the measurement basis) are possible. A connection to
the quantum Zeno effect is briefly discussed and two specific
examples are presented to demonstrate the proposed methods. The main
feature of our methods is their robustness to uncertainties in
system Hamiltonians. The sliding mode control method has potential
applications to preparation of quantum states and quantum error
correction.

This paper is organized as follows. Section 2 defines two classes of
candidate sliding modes and formulates the two quantum control
problems considered in this paper. In Section 3 we present a sliding
mode control method combining time-optimal control design and
periodic projective measurements for quantum systems where an
eigenstate is identified as a sliding mode. An illustrative example
of a two-level quantum system is analyzed in detail. Section 4
proposes a control design method using quantum amplitude
amplification and periodic projective measurements for controlling
quantum systems when the sliding mode is a state subspace. An
example of a three-level system is presented to demonstrate the
proposed method. Concluding remarks are given in Section 5.

\section{Sliding modes and problem formulation}
In this paper, we consider finite dimensional ($N$-level) quantum
systems whose dynamic evolution is described as follows (setting
$\hbar=1$ in this paper)
\begin{equation}\label{controlled model}
i\frac{\partial}{\partial t}|\psi(t)\rangle
=H_{0}|\psi(t)\rangle+\sum_{k}u_{k}(t)H_{k}|\psi(t)\rangle ; \ \ \ \
\ \ \ \ |\psi(t=0)\rangle=|\psi_{0}\rangle,
\end{equation}
where $H_{0}$ is the free Hamiltonian, $u_{k}(t)\in \mathbf{R}$ is
the external control, and $\{H_{k}\}$ is a set of time-independent
Hamiltonian operators. If the eigenvalues and corresponding
eigenstates of $H_{0}$ are denoted as $\lambda_{j}$ and
$|\phi_{j}\rangle (j=1,2,\dots,N)$ (i.e.,
$H_{0}|\phi_{j}\rangle=\lambda_{j}|\phi_{j}\rangle$), respectively,
$|\psi(t)\rangle$ can be expanded as:
\begin{equation}\label{superpositionstate}
|\psi(t)\rangle=\sum_{j=1}^{N}c_{j}(t)|\phi_{j}\rangle .
\end{equation}

In (\ref{controlled model}), $H=H_{0}+\sum_{k}u_{k}(t)H_{k}$ is
Hermitian which ensures that the transition from a pure state
$|\psi(0)\rangle$ to another $|\psi(t)\rangle$ can be accomplished
through a unitary transformation $U(t)$; i.e.,
$|\psi(t)\rangle=U(t)|\psi(0)\rangle$. The control problem is
converted into the problem that given an initial state and a target
state, find a set of controls $\{u_{k}(t)\}$ in (\ref{controlled
model}) or a unitary transformation $U(t)$ to drive the controlled
system from the initial state into the target state.

\subsection{Sliding modes}
Sliding mode control (SMC) is a useful approach to robust controller
design for electromechanical systems \cite{Utkin et al 1999}. In
SMC, a sliding surface (sliding mode) and a switching function are
determined \cite{Edwards and Spurgeon 1998}. A sliding mode is
generally defined as a specified state region where the system has
desired dynamic behavior. A switching function is designed to
control the system state to the sliding mode. SMC has two main
advantages \cite{Edwards and Spurgeon 1998}: (i) the dynamic
behavior of the system may be determined by the particular choice of
switching function; (ii) the closed-loop response becomes totally
insensitive to a particular class of uncertainties. It is the order
reduction property and low sensitivity to uncertainty that makes SMC
an efficient tool for controlling complex high-order systems subject
to uncertainties \cite{Utkin et al 1999}.

To apply the idea of SMC to quantum systems, we first need to define
a sliding mode where the quantum system has desired dynamics. A
sliding mode can be represented as a functional of the state
$|\psi\rangle$ and Hamiltonian $H$; i.e., $S(|\psi\rangle, H)=0$.
For example, an eigenstate $|\phi_{j}\rangle$ of $H_{0}$ can be
selected as a sliding surface. In this case, we can define
$S(|\psi\rangle, H)=1-|\langle \psi|\phi_{j}\rangle|^{2}$. If the
initial state $|\psi_{0}\rangle$ is in the sliding mode; i.e.,
$S(|\psi_{0}\rangle, H)=1-|\langle \psi_{0}|\phi_{j}\rangle|^{2}=0$,
we can easily prove that the quantum system will maintain its state
in this surface under the action of the free Hamiltonian $H_{0}$
only. In fact, $|\psi(t)\rangle=e^{-iH_{0}t}|\psi_{0}\rangle$ and
$$S(|\psi(t)\rangle,H)=1-|\langle
\psi(t)|\phi_{j}\rangle|^{2}=1-|\langle
\psi_{0}|e^{iH_{0}t}|\phi_{j}\rangle|^{2}=1-|\langle
\psi_{0}|\phi_{j}\rangle e^{i\lambda_{j}t}|^{2}$$
$$=1-|\langle \psi_{0}|\phi_{j}\rangle|^{2}
|e^{i\lambda_{j}t}|^{2}=0. \ \ \ \ \ \ \ \ \ \ \ \ \ \ \ \ \ \ \ \ \
\ \ \ \ \ $$ That is, an eigenstate of $H_{0}$ can be identified as
a sliding mode.

More generally, an invariant state subspace of a quantum system
satisfying $S(|\psi\rangle, H)=0$ can be defined as a sliding mode.
For example, the wavefunction controllable subspace considered in
\cite{Dong and Petersen 2008} can be identified as a sliding mode.
Considering a simple control $u(t)$ (i.e., $k=1$ in (\ref{controlled
model})) and substituting (\ref{superpositionstate}) into
(\ref{controlled model}), we denote $C(t)=\{c_{1}(t), c_{2}(t),
\dots, c_{N}(t)\}$ and can obtain \cite{Turinici and Rabitz 2003},
\cite{Turinici and Rabitz 2001}:
\begin{equation}\label{model}
i\dot{C}(t)=AC(t)+u(t)BC(t); \ \ \ \ \ \ C(t=0)=C_{0};
\end{equation}
\begin{equation}\label{coefficient}
C_{0}=(c_{0j})_{j=1}^{N}; \ \ \ \
c_{0j}=\langle\phi_{j}|\psi_{0}\rangle, \ \ \ \ \ \
\sum_{j=1}^{N}|c_{0j}|^{2}=1,
\end{equation}
where $A$ and $B$ correspond to the operators $H_{0}$ and $H_{1}$
(i.e., $k=1$ in (\ref{controlled model})), respectively. Consider
the Model I presented in \cite{Dong and Petersen 2008},
\begin{equation}\label{model1}
i\dot{C}(t)=AC(t)+u(t)BC(t)=(A+u(t)B)C(t); \ \ \ \ \ \ C(t=0)=C_{0},
\end{equation}
where the free Hamiltonian of the five-level system is
$A=\text{diag}\{1.0, 1.2, 1.3, 2.0, 2.15\}$ \cite{Tersigni et al
1990} and the control Hamiltonian $H_{u}=u(t)B$ is as follows:
\begin{equation}
H_{u}=
\begin{pmatrix}
  0   & 0  & 0  & u(t)  & u(t)   \\
  0   & 0  & 0  & 0  & 0   \\
  0   & 0  & 0  & 0  & 0   \\
  u(t)   & 0  & 0  & 0  & u(t)   \\
  u(t)   & 0  & 0  & u(t)  & 0  \\
\end{pmatrix}.
\end{equation}

In \cite{Dong and Petersen 2008}, we have proven that the subspace
$\Omega$ spanned by
$\{|\phi_1\rangle,|\phi_4\rangle,|\phi_5\rangle\}$ is a wavefunction
controllable subspace \cite{Dong et al 2008-1}. We may select
$\Omega$ as a sliding surface. We can easily prove that if the
initial state of this system is in this sliding surface, its state
will be maintained in this surface under the action of Hamiltonian
$H=A+u(t)B$. In fact, we can express the sliding mode as follows:
$$S(|\psi\rangle, H)=1-(|\langle \psi|\phi_{1}\rangle|^{2}+|\langle \psi|\phi_{4}\rangle|^{2}
+|\langle \psi|\phi_{5}\rangle|^{2})=0$$ If $S(|\psi_{0}\rangle,
H)=0$, we can obtain $S(|\psi(t)\rangle, H)=0$. From
$S(|\psi_{0}\rangle, H)=0$, it is clear that we can obtain
$|c_{01}|^{2}+|c_{04}|^{2}+|c_{05}|^{2}=1$. Now we obtain the
following equation from (\ref{model1}):
\begin{equation}\label{sixequations}
\left(%
\begin{array}{c}
  \dot{c}_{1}(t) \\
  \dot{c}_{4}(t) \\
  \dot{c}_{5}(t) \\
\end{array}%
\right)
=\left(%
\begin{array}{ccc}
  -i & -iu(t) & -iu(t) \\
  -iu(t) & -2.0i  & -iu(t) \\
  -iu(t) & -iu(t) & -2.15i \\
\end{array}%
\right) \left(%
\begin{array}{c}
  c_{1}(t) \\
  c_{4}(t) \\
  c_{5}(t) \\
\end{array}%
\right).
\end{equation}
After straightforward calculations we obtain the following
relationship :
\begin{equation}
\frac{d}{dt}(|c_{1}(t)|^{2}+|c_{4}(t)|^{2}+|c_{5}(t)|^{2})=0 .
\end{equation}
From this, it is clear that
$|c_{1}(t)|^{2}+|c_{4}(t)|^{2}+|c_{5}(t)|^{2}=|c_{01}|^{2}+|c_{04}|^{2}+|c_{05}|^{2}=1$.
That is, $S(|\psi(t)\rangle, H)=0$; i.e., the state of this system
will be maintained in this surface under the action of Hamiltonian
$H=A+u(t)B$.

Some other state subspaces such as decoherence-free subspace
\cite{Lidar et al 1998}, \cite{Kwiat et al 2000} can also be
defined as sliding modes. However, this paper will focus on the
two particular classes of sliding modes. The first class
corresponds to a sliding mode which is an eigenstate. The second
class corresponds to a sliding mode which is a wavefunction
controllable subspace.

\subsection{Problem formulation}
In the above subsection, we have presented two candidate sliding
modes. If a quantum system state is driven into a sliding mode, the
state will be maintained in the sliding surface under the action of
some class of Hamiltonians determined by the sliding mode. However,
in practical applications, it is inevitable that there exist noises
and uncertainties in system Hamiltonians, initial states or control
fields. An important advantage of sliding mode control is its
robustness against uncertainties. Our main motivation of introducing
sliding mode control to quantum systems is to deal with these
uncertainties. In this paper, we focus on the class of uncertainties
which can be approximated as a perturbation in the system
Hamiltonian. For example, the unitary error in \cite{Pravia et al
2003} and the operational error in a quantum logic gate can be
classified into this class of uncertainties. We also suppose that
the uncertainties are bounded. We represent the uncertainties as
$H_{\Delta}=\sum_{l}\epsilon_{l}(t)H_{l}$, where $\epsilon_{l}(t)\in
\mathbf{R}$, $\sqrt{\sum_{l}\epsilon_{l}^{2}(t)}\leq \bar{\epsilon}$
($\bar{\epsilon}\in \mathbf{R^+}$) and $\{H_{l}\}$ is a set of
time-independent Hamiltonian operators. We further suppose that the
system is completely controllable~\cite{Schirmer et al 2001}.

The control problem under consideration is stated as follows: for a
given initial state, design a control law to steer the quantum
system state into and then maintain the state in a sliding mode
domain (a neighborhood containing the sliding mode) in the presence
of bounded uncertainties in the system Hamiltonian. This quantum
sliding mode control problem is greatly different from the
traditional sliding mode control. Once the uncertainties take the
state slightly away from the sliding mode, there is always a finite
probability (we call it the probability of failure) that the system
state will collapse out of the sliding mode domain when one makes a
measurement on this system. Hence, if the allowed probability of
failure is $p_{0}$, we may define the sliding mode domain
$\mathcal{D}=\{|\psi\rangle: |\langle \psi|\Phi\rangle|^{2}\geq
1-p_{0}, |\Phi\rangle \in \{|\Phi\rangle: S(|\Phi\rangle, H)=0\}
\}$. We expect that the control law can ensure that the system state
remains in the sliding mode domain $\mathcal{D}$ except that a
measurement operation may take it away from $\mathcal{D}$ with a
small probability (not greater than $p_{0}$). The quantum sliding
mode control problem includes three main subtasks: (I) for any
initial state (assumed to be known), design a control law to drive
the system state into a defined sliding mode domain $\mathcal{D}$;
(II) design a control law to maintain the system state in
$\mathcal{D}$; (III) design a control law to drive the system state
back to $\mathcal{D}$ if a measurement takes it away from
$\mathcal{D}$. For convenience, we suppose that there exist no
uncertainties during the control processes (I) and (III). In
particular, we consider the following two quantum control problems.

(1) Quantum Control Problem 1 (\textbf{QCP1}): For an uncertain
quantum system in which an eigenstate $|\phi_{j}\rangle$ of $H_{0}$
defines a sliding mode (i.e., $S(|\psi\rangle, H)=1-|\langle
\psi(t)|\phi_{j}\rangle|^{2}=0$), (I) drive the controlled quantum
system state into the sliding mode domain
$\mathcal{D}=\{|\psi\rangle: |\langle \psi|\phi_{j}\rangle|^{2}\geq
1-p_{0}\}$; (II) maintain the system state in $\mathcal{D}$ except
that measurements may take it away from $\mathcal{D}$ with at most
probability $p_{0}$; (III) if the system state is taken away from
$\mathcal{D}$, design a control law to drive it back to
$\mathcal{D}$.

(2) Quantum Control Problem 2 (\textbf{QCP2}): For an uncertain
quantum system in which a wavefunction controllable subspace
$\Omega$ defines a sliding mode (i.e., $S(|\psi\rangle,
H)=1-|\langle \psi(t)|\Phi\rangle|^{2}=0$ where $|\Phi\rangle \in
\Omega$), (I) drive the controlled quantum system state into the
sliding mode domain $\mathcal{D}=\{|\psi\rangle: |\langle
\psi|\Phi\rangle|^{2}\geq 1-p_{0}, |\Phi\rangle \in \Omega \}$; (II)
maintain the state in $\mathcal{D}$ except that measurements may
take it away from $\mathcal{D}$ with at most probability $p_{0}$;
(III) if the system state is taken away from $\mathcal{D}$, design a
control law to drive it back to $\mathcal{D}$.

\section{SMC based on time-optimal design and periodic measurements}

\subsection{The general method}
In this section, we consider \textbf{QCP1}, in which an eigenstate
is identified as a sliding mode. Our first task is to design a
control law to drive the system state to this chosen sliding
surface. Since we ignore the effect of uncertainties during the
control processes (I) and (III), we wish to accomplish this task as
quickly as possible. Hence, we will use a time-optimal control
approach for this task. For the subtask (II), we use periodic
projective measurements to achieve our goal. In coherent control,
measurement is usually regarded as having deleterious effects.
Recent results have shown that quantum measurements can be combined
with unitary transformations to complete some quantum manipulation
tasks and enhance the capability of quantum control \cite{Vilela
Mendes and Man'ko 2003}, \cite{Gong and Rice 2004}, \cite{Shuang et
al 2007}-\cite{Romano and D'Alessandro 2006-2}. For example, Vilela
Mendes and Man'ko \cite{Vilela Mendes and Man'ko 2003} showed that
nonunitarily controllable systems might become controllable by using
``measurement plus evolution". Roa \emph{et al.} \cite{Roa et al
2006} have used sequential measurements to control quantum systems.
Rabitz and coworkers \cite{Shuang et al 2007}-\cite{Dong et al
2008-3} have demonstrated that projective measurements can serve as
a control tool. In this section, we will combine time-optimal
control and periodic projective measurements to accomplish sliding
mode control of quantum systems.

The steps of the control algorithm for \textbf{QCP1} are as follows:
\begin{enumerate}
    \item
    Select an eigenstate $|\phi_{j}\rangle$ of $H_{0}$ as a sliding mode $S(|\psi\rangle, H)=0$;
    \item For a known initial state $|\psi_{0}\rangle$, design a time-optimal control law that can drive $|\psi_{0}\rangle$ to the
sliding mode $S$;
    \item For eigenstates
$|\phi_{k}\rangle$ ($k\neq j$), design corresponding time-optimal
control laws that can drive $|\phi_{k}\rangle$ to $S$ using a
similar method to that in (ii);
    \item For given $p_{0}$ and $\epsilon$, design the period $T$ for the projective measurements;
    \item Use the designed control law to drive the system state to $S$, then implement periodic projective
    measurements with the period $T$ to maintain the system state in $\mathcal{D}=\{|\psi\rangle: |\langle
\psi|\phi_{j}\rangle|^{2}\geq 1-p_{0}\}$. If the state collapses to
$|\phi_{k}\rangle$ due to a measurement, we use the
    corresponding control law to drive it to $S$ and then continue to make periodic projective measurements.
\end{enumerate}

From the above, we can see that the design of a time-optimal control
law and the selection of the period $T$ for projective measurements
are the two most important tasks in this control algorithm. The
time-optimal control problem has been an interesting topic in
quantum control, in which it is required to design a control law to
achieve a desired state transfer in a minimum time in order to
minimize the effects of relaxation and decoherence \cite{Khaneja et
al 2001}, \cite{Sugny et al 2007}. Khaneja and coworkers
\cite{Khaneja et al 2001} have studied time-optimal control of spin
systems under the assumption of unbounded controls. Boscain and
Mason \cite{Boscain and Mason 2006} have investigated the
time-optimal control problem for a spin $1/2$ system with bounded
controls. For several simple quantum systems (e.g., two-level
systems), it is possible to obtain analytical results. However, it
is generally difficult to find a complete solution for
high-dimensional quantum systems. In these cases, it may be useful
to develop a numerical simulation method to find an approximate
solution.

Another important task is to design the measurement period $T$ so
that the control law can guarantee control performance. An extreme
case occurs when $T\rightarrow 0$. That is, after the quantum system
state is driven into the sliding mode, we make frequent
measurements. This corresponds to the quantum Zeno effect
\cite{Itano et al 1990} which can guarantee that the state is
maintained in the sliding mode in spite of the existence of
uncertainties. However, it is a difficult task to make such frequent
measurements in practical applications. We may think that the
smaller $T$ is, the bigger the cost of measurements becomes. Hence,
we wish to design a period $T$ which is as large as possible, when
we have a bound $\bar{\epsilon}$ on the uncertainties and require a
probability of failure $p_{0}$. In the following subsection, we will
present a specific example of a two-level quantum system to
demonstrate how to design the period $T$.

\newtheorem{remark}{Remark}
\begin{remark}
It is clear that the sliding mode control approach can be used in
the preparation and protection of quantum states under uncertainty
conditions. Moreover, if the initial state is unknown, we can first
make a projective measurement. If the result is the eigenstate
$|\phi_j\rangle$, we would continue to implement periodic projective
measurements to maintain the system state in $\mathcal{D}$.
Otherwise, we would use a corresponding control law to drive the
system state to $S$. This slight amendment to our approach enables
our method to achieve robustness against variations in the initial
state as well as robustness to uncertainties in the system
Hamiltonian.
\end{remark}

\subsection{An illustrative example: two-level system}
To demonstrate the proposed method, here we consider a two-level
quantum system, which can be used as a quantum bit (qubit) and has
important potential applications in quantum information. In
practical applications, we often use the density operator $\rho$ to
describe the state of a quantum system. For a pure state
$|\psi\rangle$, the corresponding density operator is $\rho\equiv
|\psi\rangle \langle \psi|$. For a two-level quantum system, the
state $\rho$ can be represented in terms of the Bloch vector
$\mathbf{r}=(x,y,z)=(\text{tr}\{\rho\sigma_{x}\},\text{tr}\{\rho\sigma_{y}\},\text{tr}\{\rho\sigma_{z}\})$:
\begin{equation}\label{eq14}
\rho=\frac{1}{2}(I+\mathbf{r}\cdot \sigma) .
\end{equation}
where $\sigma=(\sigma_{x},\sigma_{y},\sigma_{z})$ are the Pauli
matrices described as follows:
\begin{equation}
\sigma_{x}=\begin{pmatrix}
  0 & 1  \\
  1 & 0  \\
\end{pmatrix}, \ \ \ \
\sigma_{y}=\begin{pmatrix}
  0 & -i  \\
  i & 0  \\
\end{pmatrix}, \ \ \ \
\sigma_{z}=\begin{pmatrix}
  1 & 0  \\
  0 & -1  \\
\end{pmatrix}.
\end{equation}
The evolution of $\rho$ can be described by the equation
\begin{equation}
\dot\rho=-i[H, \rho],
\end{equation}
where $H=H_{0}+\sum_{k=x,y,z}u_{k}(t)I_{k}$,
$H_{0}=I_{z}=\frac{1}{2}\sigma_{z}$, $I_{x}=\frac{1}{2}\sigma_{x}$
and $I_{y}=\frac{1}{2}\sigma_{y}$.

Without loss of generality, we select the sliding mode as
$S(|\psi\rangle, H)=1-|\langle \psi|0\rangle|^2=0$. This means that
we select the eigenstate $|0\rangle$ of $H_{0}$ as the sliding mode.
If we have driven the system state to the sliding mode at time
$t_{0}$, it will be maintained in this sliding mode using only the
free Hamiltonian $H_{0}$; i.e., $S(|\psi_{(t\geq t_{0})}\rangle,
H_{0})\equiv 0$. We assume that the possible uncertainties in the
system Hamiltonian are represented by
$H_{\Delta}=\epsilon_{x}(t)I_{x}+\epsilon_{z}(t)I_{z}$, where
$\sqrt{\epsilon_{x}^{2}(t)+\epsilon_{z}^{2}(t)}\leq \bar{\epsilon}$.
Now we use the control algorithm in Section 3.1 to accomplish the
robust control design.

For simplicity, we assume $|\psi_{0}\rangle=|1\rangle$ (this makes
the subtasks (I) and (III) in \textbf{QCP1} become the same) and
consider $H_{u}=u(t)I_{x}$. If $u(t)$ is not bounded, it is
convenient to design a time-optimal control law to drive $|1\rangle$
to the sliding mode $|0\rangle$ using the results in \cite{Khaneja
et al 2001}, \cite{Dong IET 2009}. From Theorem 1 in \cite{Dong IET
2009}, we learn that the minimum time required to accomplish this
task is $0$. If $u(t)$ is bounded (i.e., $|u(t)|\leq V$, $V\in
\mathbf{R^+}$), we can use the method in \cite{Boscain and Mason
2006} to design a time-optimal control law to drive $|1\rangle$ to
$|0\rangle$.

Now we consider the design of the measurement period $T$. First
consider an uncertainty represented by
$H_{\Delta}=\epsilon_{z}(t)I_{z}$ (where $|\epsilon_{z}(t)|\leq
\bar{\epsilon}$). If $S(|\psi_{0}\rangle, H)=0$, for
$H_{\Delta}=\epsilon_{z}I_{z}$ we have
$$S(|\psi(t),H)=1-|\langle \psi(t)|0\rangle|^{2}=1-|\langle
\psi_{0}|e^{i(H_{0}+\epsilon_{z}I_{z})t}|0\rangle|^{2}=1-|\langle
\psi_{0}|0\rangle|^{2}
|e^{i(\frac{1}{2}+\frac{1}{2}\epsilon_{z})t}|^{2}=0 .
$$
This uncertainty does not take the system state away from the
sliding mode. Hence, we may ignore this uncertainty in the following
analysis.

Now consider a bit-flip type uncertainty of the form
$\epsilon(t)I_{x}$ ( $|\epsilon(t)|\leq \bar{\epsilon}$). We have
the following theorem.

\begin{theorem}
For a two-level quantum system with the initial state
$(x_{0},y_{0},z_{0})=(0,0,1)$ (i.e., $|0\rangle$), the system
evolves to $(x^{a}_{t},y^{a}_{t},z^{a}_{t})$ and
$(x^{b}_{t},y^{b}_{t},z^{b}_{t})$ under the action of the
Hamiltonians $H^{a}=I_{z}+\epsilon(t)I_{x}$ (where
$|\epsilon(t)|\leq \bar{\epsilon}$) and
$H^{b}=I_{z}+\tilde{\epsilon}I_{x}$ (where $\tilde{\epsilon}=
+\bar{\epsilon}$ or $-\bar{\epsilon}$), respectively. Then for
arbitrary $t\in [0, \frac{\pi}{\sqrt{1+\bar{\epsilon}^{2}}}]$,
$z^{a}_{t}\geq z^{b}_{t}$.
\end{theorem}

The proof of this theorem is presented in Appendix. Theorem 1 shows
that $z_{t}^{b}$ can be taken as an estimate of the bound on
$z_{t}^{a}$. From the proof, it follows that when
$t=\frac{\pi}{\sqrt{1+\bar{\epsilon}^{2}}}$, the probability of
failure is $p'=\frac{\bar{\epsilon}^{2}}{1+\bar{\epsilon}^{2}}$. If
$p_{0}\leq p'$, using (\ref{theorem1-bound-solution}) in Appendix we
can choose $T$ according to the following relationship:
\begin{equation}\label{period1}
p_{0}=\frac{1-z_{T}}{2}=\frac{\bar{\epsilon}^{2}}{1+\bar{\epsilon}^{2}}\frac{{1-\cos\sqrt{1+\bar{\epsilon}^{2}}
T}}{2} .
\end{equation}

Hence, we may choose the measurement period $T$ as follows:
\begin{equation}\label{period}
T=
\begin{cases}
    \frac{1}{\sqrt{1+\bar{\epsilon}^{2}}}\arccos[1-\frac{2p_{0}(1+\bar{\epsilon}^{2})}{\bar{\epsilon}^{2}}] & \text{if}\ p_{0}\leq \frac{\bar{\epsilon}^{2}}{1+\bar{\epsilon}^{2}}; \\
    \frac{\pi}{\sqrt{1+\bar{\epsilon}^{2}}} & \text{otherwise}.
\end{cases}
\end{equation}

In quantum computation, an important result is the fact that
arbitrarily accurate quantum computation is possible provided that
the error per operation is below a threshold value \cite{Knill et al
1998}. In this application the uncertainties under consideration may
come from quantum gate errors. If we define the gate fidelity as
follows \cite{Nielsen and Chuang 2000}, \cite{Wesenberg}:
$$\mathcal{F}(U_{0}, U)=\min_{|\psi\rangle}|\langle \psi|U_{0}^{\dagger}U|\psi\rangle|,$$
a straightforward calculation shows that the gate fidelity is not
less than $\frac{1}{\sqrt{1+\bar{\epsilon}^{2}}}$ under our control
strategy. If we define the quantum gate error as
$G_{e}=1-\mathcal{F}(U_{0}, U)$, the gate error is not greater than
$1-\frac{1}{\sqrt{1+\bar{\epsilon}^{2}}}$. Hence, the proposed
method can be used in the design of robust quantum gates.

Now we consider a specific case. Suppose that the control is bounded
$|u(t)|\leq 10$, the bound of uncertainties is $\bar{\epsilon}=0.1$,
and the allowed probability of failure is $p_{0}=1.00\%$. According
to \cite{Boscain and Mason 2006}, the time-optimal control from
$|1\rangle$ to $|0\rangle$ is bang-bang control and the number of
switchings required is 1. A straightforward calculation using the
method in \cite{Boscain and Mason 2006} leads us to the conclusion
that we use $u(t)=-10$ in $t\in [0, 0.1573]$ and use $u(t)=10$ in
$t\in (0.1573, 0.3146]$. Since $p_{0}>
\frac{\bar{\epsilon}^{2}}{1+\bar{\epsilon}^{2}}$, using
(\ref{period}) we can obtain the measurement period $T=3.1260$. The
corresponding maximum gate error is not greater than
$G_{e}=1-\frac{1}{\sqrt{1.01}}=0.50\%$.

\section{SMC based on amplitude amplification and periodic
measurements}

In this section, we identify a state subspace as a sliding mode and
employ the quantum amplitude amplification method to design the
control laws. We will first introduce the quantum amplitude
amplification method, then present the control algorithm, and
finally give an illustrative example.

\subsection{Amplitude amplification}
The quantum amplitude amplification method is a powerful approach
used in many quantum algorithms~\cite{Brassard et al
1998}-\cite{Long 2001}. The central task in quantum amplitude
amplification is to find a suitable operator $\mathbf{Q}$ whose
repeated action on the initial state can increase the probability of
chosen eigenstates. If we denote $\mathbb{X}=\{|0\rangle, \dots,
|x\rangle, \dots, |N-1\rangle\}$ as a set of orthonormal basis in
the $N$-dimensional complex Hilbert space $\mathcal{H}$, a pure
state $|\psi\rangle$ of an $N$-level quantum system can be
represented as $|\psi\rangle=\sum_{x=0}^{N-1}c_{x}|x\rangle$, where
$\sum_{x=0}^{N-1}|c_{x}|^{2}=1$. A Boolean function
$\chi:\mathbb{X}\rightarrow\{0,1\}$ defines two orthogonal subspaces
of $\mathcal{H}$: the ``good" subspace and the ``bad" subspace. The
good subspace is spanned by the set of basis states $|x\rangle\in
\mathbb{X}$ satisfying $\chi(x)=1$ and the bad subspace is its
orthogonal complement in $\mathcal{H}$. We may decompose
$|\psi\rangle$ as $|\psi\rangle=|\psi_g\rangle+|\psi_b\rangle$,
where $|\psi_g\rangle=P_g|\psi\rangle$ denotes the projection of
$|\psi\rangle$ onto the good subspace with the corresponding
projector $P_g$, and $|\psi_b\rangle=(I-P_g)|\psi\rangle$ denotes
the projection of $|\psi\rangle$ onto the bad subspace (here $I$ is
the identity matrix). It is clear that the occurrence probabilities
of a ``good" state $|x\rangle$ [$\chi(x)=1$] and a ``bad" state
$|x\rangle$ [$\chi(x)=0$] upon measuring $|\psi\rangle$ are
$g=\langle\psi_g|\psi_g\rangle$ and
$b=\langle\psi_b|\psi_b\rangle=1-g$, respectively.

Let $|\psi\rangle=\mathcal{U}|0\rangle$. Given two angles $0\leq
\varphi_{1}, \varphi_{2} \leq \pi$, quantum amplitude amplification
can be realized by the following operator \cite{Brassard et al 1998}
\begin{equation}
\mathbf{Q}=\mathbf{Q}(\mathcal{U},\chi,\varphi_{1},\varphi_{2})
=-\mathcal{U}\mathcal{P}_{0}^{\varphi_{1}}\mathcal{U}^{-1}\mathcal{P}_{\chi}^{\varphi_{2}}
.
\end{equation}
The operators $\mathcal{P}_{0}^{\varphi_{1}}$ and
$\mathcal{P}_{\chi}^{\varphi_{2}}$ conditionally change the phases
of state $|0\rangle$ and the good states, respectively
\cite{Brassard et al 1998}, and they can be expressed as \cite{Dong
et al 2008-3}:
\begin{eqnarray}
\mathcal{P}_{0}^{\varphi_{1}}&=&
I-(1-e^{i\varphi_{1}})|0\rangle\langle 0| \ ,\\
\mathcal{P}_{\chi}^{\varphi_{2}}&=&
I-(1-e^{i\varphi_{2}})\sum_{\chi(x)=1}|x\rangle\langle x|.
\end{eqnarray}
The action of $\mathbf{Q}$ can be described by the following
relationship \cite{Dong et al 2008-3}:
\begin{equation}
\begin{array}{ll}
\mathbf{Q}|\psi\rangle
=&[(1-e^{i\varphi_{1}})(1-g+ge^{i\varphi_{2}})-e^{i\varphi_{2}}]|\psi_{g}\rangle\\
&+[g(1-e^{i\varphi_{1}})(e^{i\varphi_{2}}-1)-e^{i\varphi_{1}}]|\psi_{b}\rangle.
\end{array}
\end{equation}
Thus, we can amplify (or shrink) the amplitude of $|\psi_{g}\rangle$
(or $|\psi_{b}\rangle$) by a suitable selection of the parameters
$\varphi_{1}$, $\varphi_{2}$ in $\mathbf{Q}$.

\subsection{The control algorithm}
The main steps in sliding mode control based on amplitude
amplification and periodic projective measurements for \textbf{QCP2}
are as follows:
\begin{enumerate}
    \item
    Select a state subspace $\Omega$ as a sliding mode $S(|\psi\rangle, H)=0$;
    \item For a known initial state $|\psi_{0}\rangle$, identify $\Omega$ as a ``good"
    subspace and construct an amplitude amplification operator $\mathbf{Q}(|\psi_{0}\rangle,
S)$ to amplify the probability of projecting $|\psi_{0}\rangle$ into
$\Omega$;
    \item Using the probability $p_{0}$ and $\mathbf{Q}(|\psi_{0}\rangle,
S)$, determine a number $L_{0}$ of iterations required to guarantee
that the control law drives the system state into the sliding mode
domain $\mathcal{D}=\{|\psi\rangle: |\langle
\psi|\Phi\rangle|^{2}\geq 1-p_{0}, |\Phi\rangle \in \Omega \}$;
    \item For other eigenstates
$|\phi_{k}\rangle$ not in the ``good" subspace, first apply a
unitary transformation $\mathcal{U}_{k}$ on $|\phi_{k}\rangle$ to
obtain a superposition state $|\psi_{k}\rangle$. Then construct a
corresponding amplitude amplification operator
$\mathbf{Q}(|\psi_{k}\rangle, S)$ and choose the required number
$L_{k}$ of iterations using a similar method as used in (ii) and
(iii);
    \item Using $p_{0}$ and $\epsilon$, choose the period $T$ for the periodic projective measurements;
    \item Use the designed control law to drive the system state into $\mathcal{D}$, then implement periodic projective
    measurements with the period $T$ to maintain the system state in $\mathcal{D}$. If a measurement result is
    $|\phi_{k}\rangle$ which is not in $\Omega$, use the
    corresponding control law in (iv) to drive the system state into $\mathcal{D}$.
\end{enumerate}

\begin{remark}
Here we use amplitude amplification as an important part of our
control algorithm for the sliding mode control of quantum systems.
This is similar to the idea in \cite{Dong et al 2008-3}. However,
\cite{Dong et al 2008-3} does not consider the issue of robustness.
Here, our goal is to develop a new method to deal with uncertainties
in the system Hamiltonian. From the above control algorithm, it is
clear that we can design different controllers offline. Hence, this
is a convenient approach to be applied to different control tasks.
\end{remark}

\begin{remark}
In the above control algorithm, the construction of amplitude
amplification operator $\mathbf{Q}$ is dependent on the initial
state; i.e., the initial state must be known. When the initial state
is an eigenstate $|\phi_{k}\rangle$, we usually need apply a unitary
transformation $\mathcal{U}_{k}$ to drive $|\phi_{k}\rangle$ to a
superposition state before constructing $\mathbf{Q}$. Here
$\mathcal{U}_{k}$ may be a small perturbation or an easily-realized
unitary transformation. If the initial state is unknown or is a
mixed state, we can construct a Kraus-map to control the initial
state to a specified pure state (for details, see, e.g., \cite{Wu et
al 2007}) and then use the proposed control algorithm to accomplish
the control task.
\end{remark}

\begin{remark}
In quantum information, quantum error correction is an important
problem. Two important paradigms for quantum error correction have
been proposed: active error correction and passive error avoidance.
The main idea for active error correction is encoding quantum
information using redundant qubits, detecting possible errors and
then correcting the error \cite{Shor 1995}, \cite{Steane 1996},
\cite{Nielsen and Chuang 2000}. In the paradigm of error avoidance,
one may encode quantum information in a decoherence-free (noiseless)
subspace \cite{Bacon et al 1999}, \cite{Lidar et al 1998},
\cite{Kwiat et al 2000}. Our control method presented above provides
a possible unified framework for the two paradigms of quantum error
correction. We may identify a decoherence-free subspace as a sliding
mode. In the sliding mode, the system is robust against some
specified uncertainties (errors). However, some other classes of
errors may break the specific symmetry in the system-environment
interaction and take the system state away from the decoherence-free
subspace. In that case, we can design a control law to drive the
system state back to the sliding mode. A schematic demonstration of
the relationships between sliding mode control (SMC) and quantum
error correction (QEC) is shown in Figure \ref{SMC2QEC}. From this
discussion it can be seen that SMC has the potential to provide a
new method for quantum error correction. The detailed development of
this idea is an area for future research.
\begin{figure}
\centering
\includegraphics[width=4.0in]{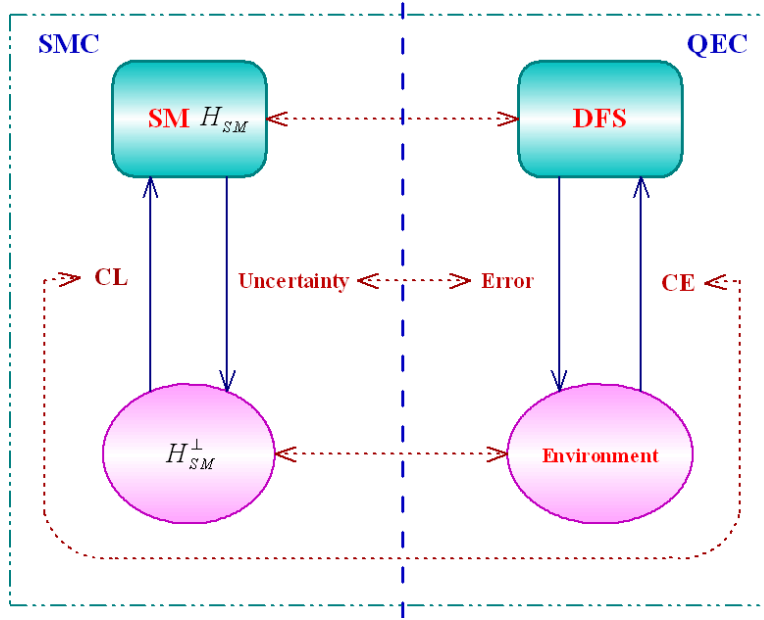}
\caption{A representation of the relationships between sliding mode
control (SMC) in \textbf{QCP2} and a unified framework of quantum
error correction (QEC). The sliding mode (SM $H_{SM}$) corresponds
to the decoherence-free subspace (DFS); the complement space
$H_{SM}^{\bot}$ corresponds to the effect of the environment in QEC;
and the uncertainties in SMC correspond to possible errors in QEC.
The subtask (I) in SMC corresponds to the initialization of QEC; the
subtask (II) in SMC corresponds to the protection of DFS in QEC; and
the control law (CL) in the subtask (III) of SMC corresponds to the
process of correcting errors (CE) in QEC.} \label{SMC2QEC}
\end{figure}

\end{remark}

\subsection{An illustrative example: three-level system}
We now consider a three-level quantum system. In (\ref{model}), let
$u(t)=1$ and $A=\text{diag}\{-1, 0, 1\}$. The matrix $B$ and the
uncertainty matrix $H_{\Delta}$ are as follows:
\begin{equation}\label{system4fig2}
B=
\begin{pmatrix}
  0   & 1  & 0     \\
  1   & 0  & 0     \\
  0   & 0  & 0     \\
\end{pmatrix}, \ \ \ \ \ \ \ \
H_{\Delta}=
\begin{pmatrix}
  0   & 0  & \epsilon(t)     \\
  0   & 0  & 0     \\
  \epsilon(t)   & 0  & 0     \\
\end{pmatrix},
\end{equation}
where $|\epsilon(t)|\leq \bar{\epsilon}$. This leads to the
following equation of the form (\ref{model})
\begin{equation}\label{exm2eq1}
i \left(%
\begin{array}{c}
  \dot{c}_{1}(t) \\
  \dot{c}_{2}(t) \\
  \dot{c}_{3}(t) \\
\end{array}%
\right)
=\left(%
\begin{array}{ccc}
  -1 & 1 & \epsilon(t) \\
  1 & 0  & 0 \\
  \epsilon(t) & 0 & 1 \\
\end{array}%
\right) \left(%
\begin{array}{c}
  c_{1}(t) \\
  c_{2}(t) \\
  c_{3}(t) \\
\end{array}%
\right).
\end{equation}

\begin{figure}
\centering
\includegraphics[width=2.0in]{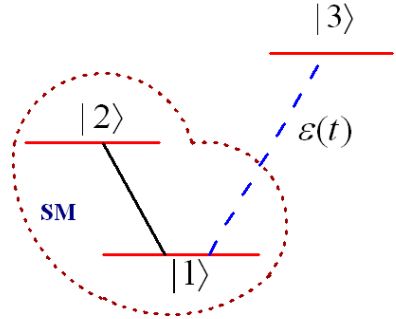}
\caption{A three-level system. The subspace spanned by $\{|1\rangle,
|2\rangle \}$ corresponds to the sliding mode (SM). The uncertainty
$\epsilon(t)$ exists between $|1\rangle$ and $|3\rangle$.}
\label{3level}
\end{figure}

The three eigenstates $\{|\phi_{1}\rangle, |\phi_{2}\rangle,
|\phi_{3}\rangle\}$ are denoted as $\{|1\rangle, |2\rangle,
|3\rangle\}$, respectively. Suppose that the initial state is
$|1\rangle$; i.e., $\{c_{1}(0), c_{2}(0), c_{3}(0)\}=\{1, 0, 0\}$.
The three-level system corresponding to (\ref{system4fig2}) can be
represented schematically as shown in Figure \ref{3level}. The form
of the $B$ matrix implies that there is a direct coupling between
states $|1\rangle$ and $|2\rangle$. The form of the uncertainty
matrix $H_{\Delta}$ implies that a disturbance coupling
$\epsilon(t)$ exists between $|1\rangle$ and $|3\rangle$. It is easy
to check using the method in Section 2.1 that the subspace spanned
by $\{|1\rangle, |2\rangle\}$ can be used as a sliding mode; i.e.,
$S(|\psi\rangle,
H)=1-(|\langle\psi|1\rangle|^{2}+|\langle\psi|2\rangle|^{2})=0$.
Hence, our control problem may correspond to a practical problem in
quantum information. For example, the two-level system with levels
$|1\rangle$ and $|2\rangle$ can be used a qubit, and the uncertainty
matrix $H_{\Delta}$ may represent a possible leakage into states
outside the qubit subspace \cite{Devitt et al 2007}.

We now consider the design of period $T$ for the periodic projective
measurements. Let $c_{j}(t)=x_{j}(t)+iy_{j}(t) (j=1,2,3)$, where
$x_{j}(t),y_{j}(t)\in \mathbf{R}$. Also, let the vector $\eta(t)$
denote
$\eta(t)=(x_{1}(t),y_{1}(t),x_{2}(t),y_{2}(t),x_{3}(t),y_{3}(t))^{T}$.
From (\ref{exm2eq1}), we obtain
\begin{equation}
\dot{\eta}(t)=F(\epsilon(t))\eta(t)
\end{equation}
where
\begin{equation}\label{exm2eq2}
F(\epsilon(t))
=\left(%
\begin{array}{cccccc}
  0 & -1 & 0 & 1 & 0 & \epsilon(t) \\
  1 & 0  & -1 & 0 & -\epsilon(t) & 0 \\
  0 & 1 & 0 &  & 0 & 0 \\
  -1 & 0  & 0 & 0 & 0 & 0 \\
  0 & \epsilon(t) & 0 & 0 & 0 & 1 \\
  -\epsilon(t) & 0  & 0 & 0 & -1 & 0 \\
\end{array}%
\right) ,
\end{equation}
and $\eta(0)=(x_{1}(0), y_{1}(0), x_{2}(0), y_{2}(0),x_{3}(0),
y_{3}(0))^{T}=(1, 0, 0, 0, 0, 0)^{T}$.

We now introduce the co-state vector
$\mathbf{\lambda}(t)=(\lambda_{1}(t), \lambda_{2}(t),
\lambda_{3}(t), \lambda_{4}(t),$ $\lambda_{5}(t),
\lambda_{6}(t))^{T}$, and obtain the corresponding Hamiltonian
function as follows:
\begin{equation}
\mathbb{H}({\eta(t),\epsilon(t),\mathbf{\lambda}(t),t})=
\lambda^{T}(t)F(\epsilon(t))\eta(t)=\epsilon(t)M(t)+N(t),
\end{equation}
where $M(t)=\lambda_{5}(t)y_{1}(t)
-\lambda_{6}(t)x_{1}(t)-\lambda_{2}(t)x_{3}(t)+\lambda_{1}(t)y_{3}(t)$
and
$N(t)=\lambda_{2}(t)x_{1}(t)-\lambda_{4}(t)x_{1}(t)-\lambda_{1}(t)y_{1}(t)
+\lambda_{3}(t)y_{1}(t)-\lambda_{2}(t)x_{2}(t)+\lambda_{1}(t)y_{2}(t)
-\lambda_{6}(t)x_{3}(t)+\lambda_{5}(t)y_{3}(t)$.
According to Pontryagin's minimum principle \cite{Kirk 1970}, a
necessary condition for $\tilde{\epsilon}(t)$ to minimize the
functional $J(\epsilon)=x_{3}^{2}(t_{f})+y_{3}^{2}(t_{f})$ is
\begin{equation}
\mathbb{H}({\tilde{\eta}(t),\tilde{\epsilon}(t),\tilde{\mathbf{\lambda}}(t),t})\leq
\mathbb{H}({\tilde{\eta}(t),\epsilon(t),\tilde{\mathbf{\lambda}}(t),t})
.
\end{equation}

Hence,
\begin{equation}\label{exm2eq3}
\tilde{\epsilon}(t)=-\bar{\epsilon} \text{sgn}M(t) .
\end{equation}
That is, the optimal control is a bang-bang control strategy; i.e.,
$\tilde{\epsilon}(t)=\tilde{\epsilon}=+\bar{\epsilon} \ \text{or}
-\bar{\epsilon}$.

Without loss of generality, now we let
$\tilde{\epsilon}(t)=\bar{\epsilon}$ and  focus on
\begin{equation}\label{exm2eq2}
\dot{\eta}(t)=F(\bar{\epsilon})\eta(t),
\end{equation}
where $\eta(0)=(1, 0, 0, 0, 0, 0)^{T}$.

Consider the optimal control with a fixed final time $t_{f}$ and a
free final state
$\eta(t_{f})=(x_{1}(t_f),y_{1}(t_f),x_{2}(t_f),y_{2}(t_f),
x_{3}(t_f),y_{3}(t_f))^{T}$. Let $J(t)=x_{3}^{2}(t)+y_{3}^{2}(t)$.
According to Pontryagin's minimum principle,
$\tilde{\lambda}(t_f)=\frac{\partial}{\partial
\eta}J(t)|_{t=t_{f}}$. From this, we obtain
$(\lambda_{1}(t_f),\lambda_{2}(t_f),\lambda_{3}(t_f),\lambda_{4}(t_f),\lambda_{5}(t_f),\lambda_{6}(t_f))
=(0,0,0,0,2x_{3}(t_{f}),2y_{3}(t_{f}))$. Now we consider another
necessary condition $\dot{\lambda}(t)=-\frac{\partial
\mathbb{H}({\eta(t),\epsilon(t),\mathbf{\lambda}(t),t})}{\partial
\eta}$ which leads to the following relationships:
\begin{equation}\label{exam2costate}
\dot{\mathbf{\lambda}}(t)=F(\bar{\epsilon})\lambda(t) .
\end{equation}

\begin{figure}
\centering
\includegraphics[width=3.6in]{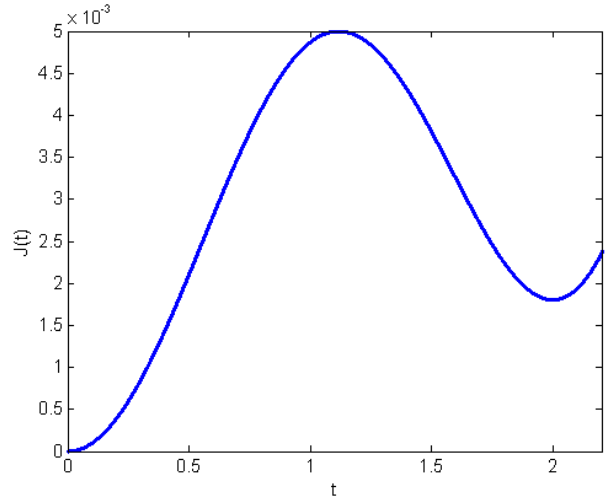}
\caption{The curve of $J(t)=x_{3}^{2}(t)+y_{3}^{2}(t)$ corresponding
to (\ref{exm2eq2}).} \label{fig3}
\end{figure}

By simulation of (\ref{exm2eq2}) and (\ref{exam2costate}), for
different values of $\bar{\epsilon}$, we find the sign of $M(t)$
does not change over the first interval for which $J(t)$ is
monotonically increasing with respect to $t$. Hence we can estimate
the required period $T$ by considering the first interval on which
$J(t)$ is monotonic. For example, if we consider
$\bar{\epsilon}=0.1$, we obtain the simulation results shown in
Figure \ref{fig3}. From this, we obtain $t_{f}=1.1160$,
$x_{3}(t_f)=-0.0055$, $y_{3}(t_f)=-0.0705$ and $J(t_f)=0.0050$.
Then we can check that $M(t)$ indeed does not change sign on $t\in
[0, t_{f}]$ as shown in Figure \ref{fig4}. Now we can design the
period $T$ using the following relationship:

\begin{figure}
\centering
\includegraphics[width=3.6in]{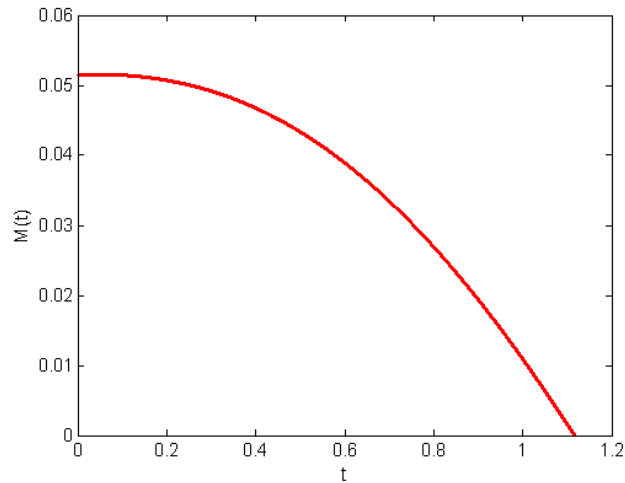}
\caption{The curve of $M(t)$ corresponding to the first monotonic
interval of $J(t)$ in Figure \ref{fig3}.} \label{fig4}
\end{figure}

\begin{equation}\label{period2}
T=
\begin{cases}
    t_{1}\ (\text{where} \ t_{1}\in [0, t_f) \ \text{and}\ J(t_{1})=p_0)\ & \text{if}\ p_{0}< J(t_f); \\
    t_f & \text{otherwise}.
\end{cases}
\end{equation}
For example, if $p_{0}=0.50\%$, we have $p_{0}\leq J(t_f)$. From
Figure \ref{fig3} and (\ref{period2}), we obtain $T=1.1160$.

Now we design a control law for the subtask (III) using the quantum
amplitude amplification method introduced in Section 4.1. For
simplicity, we use a simple choice of $\mathbf{Q}$ where
$\varphi_{1}=\pi$, $\varphi_{2}=\pi$ and the subspace spanned by
$\{|1\rangle, |2\rangle\}$ corresponds to the ``good" subspace. When
the measurement result is $|\phi_{k}\rangle=|3\rangle$, we first
apply a unitary transformation (a perturbation) $\mathcal{U}_{3}$ on
$|3\rangle$ to obtain
$|\psi_3\rangle=\mathcal{U}_{3}|3\rangle=0.0600|1\rangle+0.0800|2\rangle+0.9950|3\rangle$.
Now, using a similar method to the example in \cite{Dong et al
2008-3}, we obtain the state
$|\psi'_{3}\rangle=0.5986|1\rangle+0.7981|2\rangle+0.0682|3\rangle$
after 7 iterations of $\mathbf{Q}$. Now we make a measurement on
$|\psi'_{3}\rangle$, and the probability of failure is
$p'_{0}=0.47\%$ ($\leq 0.50\%$).

\section{Concluding remarks}
This paper focuses on a robust control problem for quantum systems
with uncertainties. The main contributions are as follows: (1) We
propose a new robust control framework based on sliding mode control
to deal with uncertainties in quantum control. Sliding mode control
is a powerful methodology in classical control theory and has many
industrial applications. Hence, the proposed method has the
potential to open a new avenue for robust control design of quantum
systems, which will provide a new useful tool for quantum
information processing with uncertainties. (2) This paper presents
two specific control algorithms for two classes of quantum control
problems. In particular, several approaches such as time-optimal
control, projective measurements and quantum amplitude amplification
are combined to design quantum control algorithms. Two examples are
also analyzed in detail using the proposed algorithms.

It should be pointed out that the results in this paper are only a
first step in developing systematic sliding mode control approaches
for quantum systems. Much further work is required and several open
problems are listed as follows: (1) The measurement period $T$
designed in this paper is relatively conservative and it may be
possible to obtain a larger value of $T$ when $p_{0}$ is relatively
large. (2) The connection between the proposed methods and quantum
error correction and quantum state preparation has been discussed.
However, the specific applications still require to be developed in
detail. Moreover, we use two model systems as the examples to
illustrate the proposed methods. A future task is to connect such
model systems to specific quantum systems (e.g., quantum optical
systems and spin systems) and to consider the experimental
implementation of the proposed methods for such real quantum
systems. (3) We only consider a simple amplitude amplification
operator $\mathbf{Q}$ in our example, and the questions of how to
determine an optimal $\mathbf{Q}$ and how to realize $\mathbf{Q}$
using specific control fields are still open problems. (4) We have
ignored the uncertainties during the control processes (I) and
(III), which require a relatively short time to accomplish the two
control objectives. An improvement worth exploring is to consider
the effect of uncertainties in all three subtasks. (5) We have
constrained the type of uncertainties to be dealt with, so future
research could be directed forwards extending the sliding mode
control approaches to deal with other types of uncertainties for
quantum systems.

\section*{Acknowledgment}
This work was supported by the Australian Research Council and was
supported in part by the National Natural Science Foundation of
China under Grant No. 60703083. The authors would like to thank two
anonymous reviewers for helpful comments. D. Dong also wishes to
thank Bo Qi, Hao Pan and Chenbin Zhang for helpful discussions.

\section*{Appendix: The proof of Theorem 1} For
$H=I_{z}+\epsilon(t)I_{x}$, using $\dot{\rho}=-i[H, \rho]$ and
(\ref{eq14}), we have
\begin{equation}
\left(%
\begin{array}{c}
  \dot{x}_{t} \\
  \dot{y}_{t} \\
  \dot{z}_{t} \\
\end{array}%
\right)
=\left(%
\begin{array}{ccc}
  0 & -1 & 0 \\
  1 & 0  & -\epsilon(t) \\
  0 & \epsilon(t) & 0 \\
\end{array}%
\right) \left(%
\begin{array}{c}
  x_{t} \\
  y_{t} \\
  z_{t} \\
\end{array}%
\right),
\end{equation}
where $(x_{0}, y_{0}, z_{0})=(0, 0, 1)$.

We now consider $\epsilon(t)$ as a control input and select the
performance measure as
\begin{equation}
J(\epsilon)=z_{t_f} .
\end{equation}
Also, we introduce the co-state vector $\lambda(t)=(\lambda_{1}(t),
\lambda_{2}(t), \lambda_{3}(t))^{T}$ and obtain the corresponding
Hamiltonian function as follows \cite{Kirk 1970}:
\begin{equation}
\mathbb{H}({\mathbf{r}(t),\epsilon(t),\mathbf{\lambda}(t),t})
=-\lambda_{1}(t)y_{t}+\lambda_{2}(t)x_{t}+\epsilon(t)(\lambda_{3}(t)y_{t}-\lambda_{2}(t)z_{t}).
\end{equation}
According to Pontryagin's minimum principle \cite{Kirk 1970}, a
necessary condition for $\tilde{\epsilon}(t)$ to minimize
$J(\epsilon)$ is
\begin{equation}
\mathbb{H}({\tilde{\mathbf{r}}(t),\tilde{\epsilon}(t),\mathbf{\tilde{\lambda}}(t),t})\leq
\mathbb{H}({\tilde{\mathbf{r}}(t),\epsilon(t),\mathbf{\tilde{\lambda}}(t),t})
.
\end{equation}
Hence the optimal control $\tilde{\epsilon}(t)$ should be chosen as
follows:
\begin{equation}\label{bangbang}
\tilde{\epsilon}(t)=-\bar{\epsilon}
\text{sgn}(\lambda_{3}(t)y_{t}-\lambda_{2}(t)z_{t}).
\end{equation}
That is, the optimal control strategy for $\epsilon(t)$ is bang-bang
control; i.e., $\tilde{\epsilon}(t)=\tilde{\epsilon}=+\bar{\epsilon}
\ \text{or} -\bar{\epsilon}$. Now we consider the system

\begin{equation}\label{eq20}
\left(%
\begin{array}{c}
  \dot{x}_{t} \\
  \dot{y}_{t} \\
  \dot{z}_{t} \\
\end{array}%
\right)
=\left(%
\begin{array}{ccc}
  0 & -1 & 0 \\
  1 & 0  & -\tilde{\epsilon} \\
  0 & \tilde{\epsilon} & 0 \\
\end{array}%
\right) \left(%
\begin{array}{c}
  x_{t} \\
  y_{t} \\
  z_{t} \\
\end{array}%
\right),
\end{equation}
where $(x_{0}, y_{0}, z_{0})=(0, 0, 1)$. Define
$\omega=\sqrt{1+\bar{\epsilon}^{2}}$.
The corresponding solution to (\ref{eq20}) is
\begin{equation}\label{theorem1-bound-solution}
\left(%
\begin{array}{c}
  x_{t} \\
  y_{t} \\
  z_{t} \\
\end{array}%
\right)
=\left(%
\begin{array}{c}
  -\frac{\tilde{\epsilon}}{1+\bar{\epsilon}^{2}}\cos\omega t+\frac{\tilde{\epsilon}}{1+\bar{\epsilon}^{2}} \\
  -\frac{\tilde{\epsilon}}{\sqrt{1+\bar{\epsilon}^{2}}}\sin\omega t \\
  \frac{\bar{\epsilon}^{2}}{1+\bar{\epsilon}^{2}}\cos\omega t+\frac{1}{1+\bar{\epsilon}^{2}} \\
\end{array}%
\right).
\end{equation}

Now consider the optimal control problem with a fixed final time
$t_{f}$ and a free final state
$\mathbf{r}_{t_f}=(x_{t_f},y_{t_f},z_{t_f})$. According to
Pontryagin's minimum principle,
$\tilde{\lambda}(t_f)=\frac{\partial}{\partial
\mathbf{r}}z_{t}|_{t=t_{f}}$. From this, it is straightforward to
verify that
$(\lambda_{1}(t_f),\lambda_{2}(t_f),\lambda_{3}(t_f))=(0,0,1)$. Now
let us consider another necessary condition
$\dot{\lambda}(t)=-\frac{\partial
\mathbb{H}({\mathbf{r}(t),\epsilon(t),\mathbf{\lambda}(t),t})}{\partial
\mathbf{r}}$ which leads to the following relationships:

\begin{equation}\label{eq24b}
\dot{\mathbf{\lambda}}(t)=\left(%
\begin{array}{c}
  \dot{\lambda}_{1}(t) \\
  \dot{\lambda}_{2}(t) \\
  \dot{\lambda}_{3}(t) \\
\end{array}%
\right)
=\left(%
\begin{array}{ccc}
  0 & -1 & 0 \\
  1 & 0  & -\tilde{\epsilon} \\
  0 & \tilde{\epsilon} & 0 \\
\end{array}%
\right) \left(%
\begin{array}{c}
  \lambda_{1}(t) \\
  \lambda_{2}(t) \\
  \lambda_{3}(t) \\
\end{array}%
\right),
\end{equation}
where
$(\lambda_{1}(t_f),\lambda_{2}(t_f),\lambda_{3}(t_f))=(0,0,1)$. The
corresponding solution to (\ref{eq24b}) is
\begin{equation}
\left(%
\begin{array}{c}
  \lambda_{1}(t) \\
  \lambda_{2}(t) \\
  \lambda_{3}(t) \\
\end{array}%
\right)
=\left(%
\begin{array}{c}
  -\frac{\tilde{\epsilon}}{1+\bar{\epsilon}^{2}}\cos\omega(t_f-t)+\frac{\tilde{\epsilon}}{1+\bar{\epsilon}^{2}} \\
  \frac{\tilde{\epsilon}}{\sqrt{1+\bar{\epsilon}^{2}}}\sin\omega(t_f-t) \\
  \frac{\bar{\epsilon}^{2}}{1+\bar{\epsilon}^{2}}\cos\omega(t_f-t)+\frac{1}{1+\bar{\epsilon}^{2}} \\
\end{array}%
\right).
\end{equation}

We obtain
\begin{equation}
\lambda_{3}(t)y_{t}-\lambda_{2}(t)z_{t}
=\frac{-\tilde{\epsilon}}{\omega^{3/2}}[\sin\omega
t+\bar{\epsilon}^{2}\sin\omega t_f +\sin\omega(t_f-t)].
\end{equation}
It is easy to show that the quantity
$(\lambda_{3}(t)y_{t}-\lambda_{2}(t)z_{t})$ occurring in
(\ref{bangbang}) does not change its sign when $t_{f}\in [0,
\frac{\pi}{\sqrt{1+\bar{\epsilon}^{2}}}]$ and $t \in [0,t_{f}]$.
That is to say, $\epsilon(t)=\tilde{\epsilon}$ is the optimal
control when $t\in [0, \frac{\pi}{\sqrt{1+\bar{\epsilon}^{2}}}]$.
Hence $z^{a}_{t}=z_{t}(\epsilon(t))\geq
z_{t}(\tilde{\epsilon})=z^{b}_{t}$.

\end{document}